\documentclass[12pt]{article}
\usepackage{amssymb,amsmath,esint,titlesec,mathrsfs}
\titleformat*{\section}{\normalsize\bf}
\titleformat*{\subsection}{\small\bf}

\begin{document}


\begin{titlepage}

\setlength{\baselineskip}{18pt}

                               \vspace*{0mm}

                             \begin{center}

{\Large\bf On the origin of escort distributions for q-entropies}

                                   \vspace{40mm}

              \large\sf  NIKOLAOS \  KALOGEROPOULOS $^\dagger$\\

                            \vspace{1mm}

  \normalsize\sf  Center for Research and Applications \\
                                  of Nonlinear Systems  \ \  (CRANS),\\
                          University of Patras, Patras 26500, Greece.\\

                            \vspace{2mm}
                         
                                    \end{center}

                            \vspace{20mm}

                     \centerline{\normalsize\bf Abstract}
                     
                           \vspace{3mm}
                     
\normalsize\rm\setlength{\baselineskip}{18pt} 

\noindent\rm We present an argument about the origin of escort distributions used in conjunction with the q-entropy in 
non-additive thermo-statistics. The origin of the escort distributions is ascribed to the fact that  the effective statistical 
description of the underlying system is provided  by the measured Gromov-Hausdorff limit of a sequence of manifolds 
having a warped product metric structure.  We interpret the entropic parameter \ q \ appearing 
in escort distributions as a ``dimension" related to the fibration structure of the underlying phase spaces.\\

                       \vfill

                                                                         
                             \vfill

\noindent\rule{12cm}{0.3mm}\\  
   \noindent   $^\dagger$ {\normalsize\rm Electronic mail: \ \ \ \  {\sf nikos.physikos@gmail.com}}\\

\end{titlepage}
 

                                                                                \newpage                 

\rm\normalsize
\setlength{\baselineskip}{18pt}

\section{Introduction} 

The q-entropy (also known as ``Tsallis entropy'') has occupied the attention of a part of the Statistical Mechanics commuity during the last 
three decades \cite{Tsallis}. Despite some progress in developing a thermodynamic formalism based on it, and the many articles which 
claim to have detected data which can be interpreted as its extremizing distributions, called the q-exponentials,
 under the familiar constraints from the conventional Statistical Mechanics, it is probably fair to say that the origin and foundations of 
 q-entropies are still largely unknown. One of the questions which have arisen 
related to the q-entropies is the origin, and the role of the entropic parameter q \cite{Alm}, \cite{4B} and of the escort distributions \cite{BS} 
which appear in several places in the developed formalism \cite{Tsallis}, \cite{Abe}, \cite{CPA}, \cite{Tan}, \cite{Liv}.\\   

In this work we attempt to provide an explanation, at the mesoscopic/macroscopic levels, for the origin and use of the escort distributions. 
We rely on the formalism of smooth metric measure spaces \cite{LV1} and our approach follows a path similar to that of \cite{NK1}, \cite{NK2}, 
but is totally distinct from \cite{NK3}.  We rely on the  results of \cite{L1}, \cite{LV2} which we interpret appropriately. 
One should turn to \cite{V} for the broader context of metric measure spaces and their Ricci curvature through optimal transport, and to \cite{WW}
for an overview of the comparison geometry  of the Bakry-\'{E}mery-Ricci tensor on which this  work relies and for all mathematical details and proofs 
which may be needed. \\  

In Section 2,  we explain the meaning and our use of warped products. In Section 3, we provide a few definitions about the measured Gromov-Hausdorff 
convergence to make this work readable to our intended audience.  Section 4 explains how the escort distributions arise from the warped product construction.
Section 5  makes some general comments on the present, and provides an outlook for future work.\\


\section{Warped products in the present context }

We will have in mind in this work the frameworks of the micro-canonical and of the canonical ensemble  of equilibrium Statistical Mechanics.  
 To begin with, let us assume that we have a system under study which is in contact with a set of thermostats. 
 Let the combined system plus thermostats be described by a manifold \ $\mathcal{M}$ \ of dimension \ $N$. \ We can see \ $\mathcal{M}$ \ 
 as the ``universe'', although not in a literal sense, and assume that it describes an isolated system. 
Someone might also want to assume that \ $\mathcal{M}$ \ is the phase space of a Hamiltonian system. 
In our metric treatment \ $\mathcal{M}$ \ is endowed with a Riemannian metric \ $\mathbf{g}_M$. \ 
Using a Riemannian metric is quite common in most physical models. The separation between the system under study and the set of thermostats
can be modelled topologically by assuming that \ $\mathcal{M}$ \ is a fibration, which is locally a direct product \ $\mathcal{B} \times \mathcal{F}$. \ 
In this product, \ $\mathcal{B}$ \ is a manifold of dimension \ $n$ \ which describes the system under study, and \ $\mathcal{F}$ \ is a manifold of dimension
\ q \ which represents each ``thermostat'' coupled to our system of study.  One could assume \ $\mathcal{M}$ \ to have non-trivial 
algebraic-topological structure, but we will not need to elaborate at this point  inasmuch it does not provide any obstructions 
and allows the constructions of interest to go through.\\

 In our previous work \cite{NK2},  we assumed that such a structure should be a Riemannian submersion.  This assumption turns out to be   
 too general and too flexible for our purposes. Instead, let us assume something more specific: 
 that the metric structures inherited by \ $\mathcal{B}$ \ which we call it a ``leaf'', and \ $\mathcal{F}$ \ which we call  a ``fiber'' in this mathematical construction, 
are inherited from \ $\mathbf{g}_\mathcal{M}$ \  through some kind of projections.
To keep some flexibility  and still allow for differences among the fibers \ $\mathcal{F}$, \ let us assume that the metric \ 
$\mathbf{g}_\mathcal{M}$ \ has a warped product form                     
\begin{equation}
       \mathbf{g}_\mathcal{M} = \mathbf{g}_\mathcal{B} + (\varepsilon e^{-f})^2 \mathbf{g}_\mathcal{F}
\end{equation}
Here \ $f: \mathcal{B} \rightarrow \mathbb{R}$ \ is assumed to be  smooth,
and let \ $\varepsilon\in\mathbb{R}$ \ be a free parameter which will be used to create a sequence, when \ $\varepsilon$ \ will be taken to 
approach zero, in what follows. \\

One can see that topologically \ $\mathcal{M} = \mathcal{B} \times \mathcal{F}$. \ Condition (1) generalizes the direct sum metric \
$\mathbf{g}_\mathcal{M} = \mathbf{g}_\mathcal{B} \oplus \mathbf{g}_\mathcal{F}$ \ through the use of the ``warping'' function 
\ $e^{-f}: \mathcal{B} \rightarrow \mathbb{R}$ \ whose role is to ``weigh'', metrically at least, each fiber differently. The ``warping'' function \ 
$e^{-f}$ \ is taken to be positive in order to maintain the Riemanian character of the metric  \ $\mathbf{g}_\mathcal{M}$ \ as well as for the 
ascertaining that  the metrics of this construction will have negative curvature in the construction of \cite{BO}.\\

One can see in (1) that the metric projections onto the leaves \ $\mathcal{B}$ \ are isometries and that the metric 
projections onto the fibers \ $\mathcal{F}$ \ are homotheties. Moreover one also observes in (1)
that the leaves and  fibers are orthogonal at their intesection points. 
One can also prove that the leaves \ $\mathcal{B}$ \ are totally geodesic submanifolds of \ $\mathcal{M}$, \ and that the fibers \ 
$\mathcal{F}$ \ are, in turn, totally umbilic submanifolds of \ $\mathcal{M}$ \ \cite{BO}, \cite{ON}.\\

 Warped products are extensively  studied structures in Geometry over at least six decades since \cite{BO}, 
 and a great deal is known about them \cite{ON}, \cite{Chen}. 
What is more relevant for our work is that such structures appear quite frequently in Physics, especially in the context of General Relativity 
and other geometric theories \cite{ON}, \cite{DB}. One can see warped products as generalizations of the Euclidean metric of the plane 
without the origin which is topologically \ $\mathbb{R}_+ \times \mathbb{S}^1$, \ where $\mathbb{S}^1$ stands for the unit radius circle, 
as can be seen in polar coordinate parametrization 
\ $ds^2 = dr^2 + r^2 d\theta^2$. \ More broadly,  warped products can be seen as  generalizations of surfaces of revolution where the 
fibers are circles. It is worth noticing that many familiar space-times such as the Schwarzschild and Friedmann-Lemaitre-Roberston-Walker can also be 
expressed as, multiple, warped products \cite{ON}, \cite{DB}. \\      


\section{Measured Gromov-Hausdoff convergence}

We now provide a few definitions needed to formulate our proposal, about the measured Gromov-Hausdorff distance and its associated
convergence. This form of convergence applies to metric measure spaces rather than to just Riemannian manifolds. Metric measure spaces have been 
a very active area of research during the last forty five years and references abound. We  point out  \cite{BBI}, \cite{G}, \cite{P}
for details and proofs which we omit altogether. \\

Let \ $\mathfrak{X}$ \ be an arbitrary metric space with distance function \ $d$ \ and let \ $A, \ B$  \ be arbitrary
 non-empty subsets of \ $\mathfrak{X}$. \ Let \ $A\subset \mathfrak{X}$ \ and let \ $U_r(A)$ \ indicate an $r$-neighborhood of \ $A$, \ namely 
\begin{equation}
     U_r(A) \ = \  \left\{  x\in \mathfrak{X}: d(x,a) < r, \ \ \ \forall \  a  \in A \right\}  
\end{equation} 
The Hausdorff distance \ $d_H (A, B)$ \ between \ $A$ \ and \ $B$ \ is defined as 
\begin{equation}
   d_H(A, B) \ = \ \inf \left\{ r\in [0, +\infty]: \  A\subset U_r (B) \  \  \mathrm{and} \  \ B \subset U_r(A) \right\}
\end{equation}
It turns out that the Hausdorff distance \ $d_H$ \ is indeed a metric on the set of all compact subsets of \ $\mathfrak{X}$. \\   

A question which arises is how to generalize this  distance between two different metric spaces 
\ $\mathfrak{Y}, \ \mathfrak{Z}$ \ which are not subsets of the same metric space  and which, in general, have different distance functions 
\ $d_\mathfrak{Y}, \ d_\mathfrak{Z}$. \ A way to proceed  is to consider the set  \ $\mathfrak{Y} \sqcup \mathfrak{Z}$ \ which is the disjoint union of 
$\mathfrak{Y}$ \ and \ $\mathfrak{Z}$ \ endowed with the distance function \ $d_{\mathfrak{Y}\sqcup\mathfrak{Z}}$ \ which is such that 
\begin{equation}
     d_{\mathfrak{Y}\sqcup\mathfrak{Z}} = \left\{
                     \begin{array}{ll}
                                  d_\mathfrak{Y}(y_1, y_2),  &   \mathrm{if} \ \  y_1, y_2 \in \mathfrak{Y}\\
                                  d_\mathfrak{Z}(z_1, z_2),  &   \mathrm{if} \ \ z_1, z_2 \in \mathfrak{Z}\\
                                  \infty,                                 &   \mathrm{otherwise}    \\              
                     \end{array}           \right. 
\end{equation}
More generally, one can consider a metric space \ $\mathfrak{X}$ \ and two subspaces \ $\mathfrak{Y}^\prime$, \ $\mathfrak{Z}^\prime$ \ isometric to
$\mathfrak{Y}$, \ $\mathfrak{Z}$\ respectively, where the distance functions of \ $\mathfrak{Y}^\prime$, \ $\mathfrak{Z}^\prime$ \ are the restrictions 
of the metric of \ $\mathfrak{X}$ \ to them. Then the Gromov-Hausdorff distance \ $d_{GH}$ \ between \ $\mathfrak{Y}$ \ and \ $\mathfrak{Z}$ \ is given by
\begin{equation}    
        d_{GH}(\mathfrak{Y}, \mathfrak{Z}) = d_H (\mathfrak{Y}^\prime, \mathfrak{Z}^\prime )   
\end{equation}
for all possible isometric embeddings of \ $\mathfrak{Y}$, \ $\mathfrak{Z}$ \ into all possible ambient spaces \ $\mathfrak{X}$.  \. 
It turns out that it is sufficient to consider in the definition of the Gromov-Hausdorff distance as ambient space \ $\mathfrak{X}$ \ the disjoint union \ 
$\mathfrak{Y}^\prime \sqcup\mathfrak{Z}^\prime$ \ with the distance function (4). \\   

It turns out that the 
Gromov-Hausdorff distance is indeed a metric on the space of isometry classes of compact metric spaces, giventhat the distance between two such isometric 
spaces is zero. The above results can be extended to non-compact metric spaces either through a pointed version of the above definitions
or by use of ultralimits, but we will not expand upon such definitions as we will only need to consider compact manifolds in this work, as will be seen below. \\
   
Given the Gromov-Hausdorff distance, one can define convergence within the set of compact metric spaces, following the familiar definition: 
a sequence of compact metric spaces \ $\mathfrak{X}_n, \ n\in\mathbb{N}$ \ converges to a compact metric space \ $\mathfrak{X}_\infty$ \ if
\begin{equation}   
   d_{GH} (\mathfrak{X}_n, \mathfrak{X}_\infty) = 0, \ \ \ \mathrm{as} \ \ n\rightarrow\infty 
\end{equation}

The Gromov-Hausdorff convergence is a purely metric concept which does not involve volumes or any other measures.
Let us recall that there is no ``natural'' concept of volume in a general metric space. In Riemannian manifolds however, such a natural choice exists, 
mainly because Riemannian manifolds are infinitesimally Euclidean spaces which admit a natural definition of volume. By contrast, immediate 
generalizations of Riemannian manifolds, such as the Finslerian manifolds which substitute the ellipsoids as unit spheres with the boundaries of convex bodies 
do not admit  a canonical volume function \cite{G}.  Since we will be working with Riemannian manifolds, we need a version of the Gromov-Hausdorff convergence
which involves volumes, or more generally ``measures'' in metric spaces. The measures \ $\mu$ \ we consider are assumed to be 
sufficiently ``nice'' (Borel, regular) as, at this level of generality, they are adequate for our purposes. \\

Let us consider a sequence of metric measure spaces \ $( \mathfrak{X}_n, d_n, \mu_n), \  n\in\mathbb{N}$. \  If one sees  functions, 
as subsets of distributions, as dual to measures with respect to integration, then one can consider the sequence of continuous functions \ 
$f_n: \mathfrak{X}_n \rightarrow \mathbb{R}$ \  converging to \ $f_\infty: \mathfrak{X}_\infty \rightarrow \mathbb{R}$. \ 
For the convergence of measures  \ $\mu_n \rightarrow \mu_\infty$ \ one demands 
\begin{equation}
  \int_{\mathfrak{X}_n} f_n \ d\mu_n \  \rightarrow \  \int_{\mathfrak{X}_\infty} f_\infty \ d\mu_\infty, \ \ \ \mathrm{as} \ \ \  n\rightarrow \infty
\end{equation}
When both (6) and (7) are satisfied then we can state that the metric measure space \ $(\mathfrak{X}_n, d_n, \mu_n)$ \ converges to 
the metric measure space \ $(\mathfrak{X}_\infty, d_\infty, \mu_\infty)$ \  in the measured Gromov-Hausdorff sense.\\


\section{Escort probabilities from warped product metrics} 

We return to the warped product metrics (1) and the main line of our argument. 
First of all, we have to state that the manifolds \ $\mathcal{M}, \ \mathcal{B}, \ \mathcal{F}$ \ whose metrics are related by (1) are assumed to be 
compact.  This is a reasonable requirement since all these three manifolds are essentially the phase spaces of the ``universe'' \ $\mathcal{M}$, \ of the 
system under study \ $\mathcal{B}$ \ and of the thermostats \ $\mathcal{F}$. \ We are working within the framework of Hamiltonian systems where the 
total energy of the ``universe'' \ $\mathcal{M}$ \ is conserved. This essentially amounts to working within each constant energy hypersurface of \ $\mathcal{M}$ \ 
which is compact, if we assume that the Hamiltonian systems we are working with involve only proper functions. This is usually the case with polynomial, 
sinusoidal/cosinusoidal, exponential etc interactions. \\

One has to be careful at this point not to trivially apply the arguments of this work to
interactions expressed through rational functions, logarithms, or higher transcendental functions which could give rise to infinites and hence to 
non-compact phase spaces or to spaces with singularties.  This does not mean 
that the final result of the current argument will not be valid, but it just indicates that one would have to model such systems differently, 
probably using the pointed measured Gromov-Hausdorff convergence and pay attention to how infinites which may emerge should be handled.     
In our case of compact \ $\mathcal{M}$, \ by projection one gets compact \ $\mathcal{B}$ \ and \ $\mathcal{F}$ \ and therefore we can apply the above
concepts of the measured Gromov-Hausdorff convergence without any problem.  \\        

One wishes to know what happens when we take the limit \ $\varepsilon \rightarrow 0$ \ in (1). \ To be more precise, we consider the measured 
Gromov-Hausdorff limit of \ $\mathcal{M}$ \ with the Riemannian distance function induced by \ $\mathbf{g}_\mathcal{M}$ \ but endowed with the 
renormalized volume  
\begin{equation}
     \widetilde{dvol_\mathcal{M}} \  = \  \frac{dvol_\mathcal{M}}{vol_\mathcal{M} (B_\mathcal{M}(1))}
\end{equation}
where \ $B_\mathcal{M}(1)$ \ indicates the ball of unit radius in \ $\mathcal{M}$. \ The reason for this volume renormalization is technical, but need not 
concern us since in the definition of probabilities and entropy where this formalism is applied, one has the freedom to appropriately renormalize 
these quantities without changing the physical content of the theory. Given this renormalized measure of \ $\mathcal{M}$, \ one finds that the 
measured Gromov-Hausdorff limit is \ $\mathcal{B}$ \ endowed with the distance induced by \ $\mathbf{g}_\mathcal{B}$ \ and with measure 
\begin{equation} 
          d\nu \ = \  (e^{-f})^q \ dvol_\mathcal{M}
\end{equation}
A way to understand this relation is as follows: the warped product (1) ``expands'' or ``contracts'' each fiber \ $\mathcal{F}$ \ by the metric factor \ 
$e^{-f}$ \ per dimension of \ $\mathcal{F}$ \ when manifold collapse is not considered. Therefore, upon collapse of the \ q \ different dimensions 
of the fiber \ $\mathcal{F}$ \ the contribution to the resulting measure is a modification of the initial measure/volume of \ $\mathcal{M}$ \ 
by an overall factor \ $(e^{-f})^q$. \\

 We would like to mention at this point, so as to also
 make the connection with the proposal of \cite{NK1} regarding the origin of the q-entropy, that this limiting process 
 gives rise to the Bakry-\'{E}mery-Ricci tensor \ $\overline{\mathrm{Ric}}_\mathcal{B}$ \ of \ $\mathcal{B}$ \ which is given by 
 \begin{equation}   
 \overline{\mathrm{Ric}}_\mathcal{B} (X,Y)  \ = \mathrm{Ric}_\mathcal{B}(X,Y) + \mathrm{Hess}_\mathcal{B} f(X,Y) - \frac{df \otimes df}{q} (X,Y) 
\end{equation}
where \ $\mathrm{Ric}_\mathcal{B}$ \ indicates the Ricci tensor of \ $\mathbf{g}_\mathcal{B}$, \ $\mathrm{Hess}_\mathcal{B}$ \ is the Hessian of 
\ $f$ \ using the Levi-Civita connection compatible with \ $\mathbf{g}_\mathcal{B}$, \ and \ $X, Y \in T\mathcal{B}$. \ As it was stated in \cite{NK1} 
lower bounds of the Bakry-\'{E}mery-Ricci tensor on a manifold are directly related to the convexity properties of the q-entropy functional on the space of 
its probability distributions with finite second moments and endowed with the Wasserstein metric \cite{LV1}, \cite{L1}, \cite{LV2}, \cite{V}.    \\  

The physical interpretation of (9) is that when the system under study \ $\mathcal{B}$ \ is coupled to the set of ``thermostats'' \ $\mathcal{F}$, \ then the combimation  
forms the ``universe'' \ $\mathcal{M}$, \ to which the microcanonical distribution with respect to \ $dvol_\mathcal{M}$ \ is applicable. We also make the tacit 
assumption about the Hamiltonian evolution on \ $\mathcal{M}$ \ being ergodic, or even better mixing, in order to follow as closely as we can the assumptions of 
Statistical Mechanics.  When one performs calculations at the level of \ $\mathcal{B}$, \ then the natural measure to use is \ $d\nu$ \ which arises 
from the microcanonical measure of \ $\mathcal{M}$, \ as seen from the above arguments and (9), which is exactly the unnormalized escort distribution \ $d\nu$ \
related to \ $dvol_\mathcal{M}$. \\
 
The reason that the escort distributions arise in this picture, is the coupling of the system \ $\mathcal{B}$ \ 
to the environment, which is the union of the ``thermostats'' \ $\mathcal{F}$ \ as in the canonical ensemble set up in equilibrium Statistical Mechanics. 
The entropic parameter \ q, \ as can be seen in (9),  is the number of the effective degrees of freedom 
entering the mesoscopic or macroscopic description of the system. This number is expressed as the dimension \ q \ of each 
``fiber''/``thermostat'' \ $\mathcal{F}$. \   We stress that the actual phase space volume of these ``thermostats'' \ $\mathcal{F}$ \ is not really relevant in this description.  
What is relevant is the number of the effective  degrees of freedom of these ``thermostats''  
which is the dimension \ q \ of \ $\mathcal{F}$. \ This conclusion is in agreement with the conclusions of \cite{4B}.\\

 One can extend the meaning of the entropic parameter \ q \ to non-integer values, by transitioning from Geometry to Analysis through the use of general 
 functions and real-valued parameters  in the place of characteristic functions of sets and integer-valued parameters \cite{V}. This is a standard procedure in 
 Analysis which allows the transition from geometric relations to analytical ones (e.g. the Brascamp-Lieb, Poincar\'{e}, Sobolev etc inequalities) 
 which extend the domain of applicability of geometric inequalities to a larger part of their parameter space.\\                  


\section{Conclusions and discussion}

In this work we tried to provide an explanation about the origin of the escort probabilities which are used extensively in the framework of non-additive,
the term non-extensive being a misnomer, Statistical Mechanics based on the q-entropy. The close relation between the origin of the escort distributions and the 
origin of the q-entropy, which was also indicated in \cite{NK1}, becomes clear. They are both part of the same 
general viewpoint according to which one starts with probabilities, the Boltzmann/Gibbs/Shannon entropy, and the microcanonical distributions of an isolated 
system, here indicated by \ $\mathcal{M}$ \ and called the ``universe'', and through a projection, whose fibers are here called ``thermostats'' \ $\mathcal{F}$, \ 
one ends with escort probabilities, the q-entropy and the canonical description of the system under study \ $\mathcal{B}$. \ Hence, there seems to be a rather 
intimate, if not necessarily unique, relation between the q-entropies and the escort distributions. \\  
                           
We believe that it is worth looking more closely into connecting the value of the entropic parameter \ q \ in q-entropies and escort distributions 
to that of the microscopic parameters or the scaling dimensions in the Hamiltonian of the microscopic system, as was suggested in \cite{4B}. 
Ideally, one should be able to derive the value of \ q, \ if one knows the microscopic dynamics, and maybe the exact coarse-graining process used to 
connect the microscopic and mesoscopic/macroscopic worlds. Whether this is feasible, 
or to what extent, remains to be seen. It would be enlightening if we could find ways to apply the above results to models of interest, to check their 
actual validity with concrete calculations, rather than leave the arguments at the current, admittedly rather abstract/theoretical, level.   \\     
  
%
%
%
%
%
%



\end{document}